\documentclass{sig-alternate}
\pdfoutput=1
\newfont{\mycrnotice}{ptmr8t at 7pt}
\newfont{\myconfname}{ptmri8t at 7pt}
\let\crnotice\mycrnotice%
\let\confname\myconfname%

\permission{Permission to make digital or hard copies of all or part of this work for personal or classroom use is granted without fee provided that copies are not made or distributed for profit or commercial advantage and that copies bear this notice and the full citation on the first page. Copyrights for components of this work owned by others than ACM must be honored. Abstracting with credit is permitted. To copy otherwise, or republish, to post on servers or to redistribute to lists, requires prior specific permission and/or a fee. Request permissions from permissions@acm.org.}
\conferenceinfo{DocEng'14,}{September 16--19, 2014, Fort Collins, Colorado, USA.}
\copyrightetc{Copyright 2014 ACM \the\acmcopyr}
\crdata{978-1-4503-2949-1/14/09\ ...\$15.00.\\
http://dx.doi.org/10.1145/2644866.2644893}
\usepackage{hyperref}
\usepackage{subfigure}
\usepackage{multicol}
\usepackage[font=normalsize,skip=0pt]{caption} 
\begin{document}
\title{The Virtual Splitter: Refactoring Web Applications for the Multiscreen Environment}
\numberofauthors{1}
\author{
\alignauthor
Mira Sarkis, Cyril Concolato, Jean-Claude Dufourd\\
\affaddr{Telecom ParisTech; Institut Mines-Telecom; CNRS LTCI}\\
\email{\{sarkis, concolato, dufourd\}@telecom-paristech.fr}
}
\maketitle

\pagenumbering{arabic}

\begin{abstract}
Creating web applications for the multiscreen environment is still a challenge. One approach is to transform existing single-screen applications but this has not been done yet automatically or generically.
This paper proposes a refactoring system. It consists of a generic and extensible mapping phase that automatically analyzes the application content based on a semantic or a visual criterion determined by the author or the user, and prepares it for the splitting process. The system then splits the application and as a result delivers two instrumented applications ready for distribution across devices. During runtime, the system uses a mirroring phase to maintain the functionality of the distributed application  and to support a dynamic splitting process. Developed as a Chrome extension, our approach is validated on several web applications, including a YouTube page and a video application from Mozilla.
\end{abstract}
\category{D.2.11}{Software Engineering}[Software Architecture]
\begin{figure*}[t!]
\centering
\includegraphics[height=1.3in,width=7in]{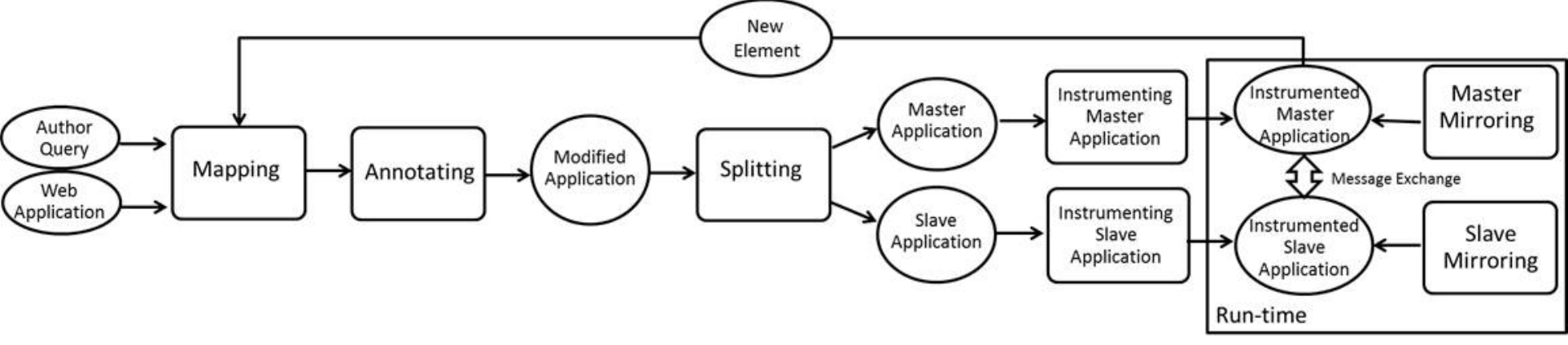} 
\caption{Virtual Splitter Architecture}
\label{algoArchitecture}
\end{figure*}
\section{Introduction}
Recently, intense research activity has been focused on multi-screen scenarios \cite{coltramEuroITV}\cite{MSS} inside home environment where cooperation between heterogeneous devices is leveraged. In such a cooperative environment, a "Multi-Screen Application" (MSA) is an application distributed across multiple connected devices, each having a screen, and designed to offer a convivial experience. Examples of MSA are: using a tablet to display additional information synchronized with a TV program, or using the interaction capabilities of smartphones (e.g., touch screen) jointly with the large screen and processing power of a PC or a TV to display media elements.

Multi-screen applications impose multiple challenges to the application developers. First, they have to design an application that leverages the multi-screen environment and copes with the diversity of devices. Then, they have to determine how the application content will be distributed across devices based on their specific capabilities, to manage the distributed content and to maintain the synchronization and consistency of the distributed content.
The use of web technologies helps reducing the complexity of these tasks and increasing the possibility of deploying a ubiquitous application.
In this paper, the term application refers to a web application.
Many applications were created before the development of the multi-screen concept. Many of them are actually made of different components that could benefit from being distributed. However, few of them were designed in a modular manner that facilitates code distribution.

With such a motivation and focusing on challenges related to authoring MSAs, this paper aims to meet one principal objective: \emph{To propose a new approach that reuses existing single-screen applications and refactors them for the multi-screen environment.} In contrast to existing works, our approach is based on mapping the application content onto available devices by automatically analyzing the content following the author or the user choices, on splitting the application into two sub-applications and synchronizing them while maintaining the overall application functionality and supporting a dynamic splitting process. 
The ability for the user to guide this splitting process is novel and opens up many usage scenarios. 

This paper is organized as follows. Section \ref{sec:SOA} compares our approach with related work. Section \ref{sec:architecture} describes the virtual splitter architecture, including the mapping, annotation, splitting and mirroring phases. An implementation of the solution, an evaluation and a survey of its limitations are given in Section \ref{sec:implementation}. 
Finally perspectives and conclusions are drawn in Section \ref{sec:conclusion}.
\section{State of the art}
\label{sec:SOA}
In prior works, a 'WebSplitter' \cite{XMLWebSplitter} was proposed to split XML-based applications, based on a metadata file. This file is unique for each application and determines which application portions, i.e., which XML elements can be seen on each user device. The splitter requires a middleware proxy that splits the application content into partial views and a client-side component 
that receives data pushed by the server. The XML splitter architecture is centralized and requires a manual mapping for each XML element of the application. 
In his research, Cheng \cite{virtualBrowser} proposed a virtual browser capable of separating the application logic from its rendering. The logic is kept within a virtual web page. Automatically the virtual browser splits the main DOM tree into multiple DOM trees and maps these trees to corresponding devices as denoted in a hint file that is specific to each application and manually created by the developer. Cross-device operations are executed in a centralized manner depending ultimately on the browser. 
Bassbouss et al.\cite{mutation-observer} outlined how to enable traditional applications to become multi-screen-ready. The application is developed as a single-screen application and requires a multi-screen enabled browser.
Based on metadata information provided manually by the developer, specific elements are assigned to a remote device while always being shown on the main device. 

In contrast to \cite{virtualBrowser} and \cite{XMLWebSplitter} our system has a decentralized architecture. Similar to \cite{mutation-observer} it delivers master-slave applications.
The common part for the three previous works is that each application is analyzed and 
mapped separately and manually by the author (via a hint file or metadata).
This means there is no generic analysis method that can be applied to a set of applications. 
On the contrary, we propose an extensible system that is capable of automatically analyzing the application and mapping its elements, thus simplifying the author's task and involving the end user in mapping her application. The author only has to determine the analysis criterion.
\section{The Virtual Splitter}
\label{sec:architecture}
\subsection{Overview}
\label{sec:overview}
A web application consists mainly of HTML, CSS and JavaScript (JS) resources, that are tightly linked.
Links exist between elements in the DOM tree (e.g., parent-child, siblings), between the DOM and JS when the JS accesses elements by specific attributes (e.g., id) or by document navigation, between the DOM and the CSS via selectors, etc. In such a context, splitting an application will break some links and cause a failure in the application look and functionality.

In addition, an application presents two dynamic aspects that make the splitting approach more complicated. On one side, splitting a web application requires support for its dynamism since elements are continuously modified, created, moved or removed during runtime. Supporting this dynamism during run-time is essential to ensure the coherence between elements in each of the distributed application.
On the other side, automatic partitioning of the application script is a hard task since JS is a flexible and dynamic language characterized by high-order functions, closures, 'eval' function which dynamically evaluates a string expression, etc.
In this paper, we take care of the links and dynamicity of the application by focusing on splitting only the HTML document, maintaining links and providing a dynamic splitting phase during run-time while keeping the JS code as a whole running on one device. The following subsections present a detailed description of the virtual splitter 
architecture as illustrated in Figure~\ref{algoArchitecture}.
\subsection{HTML Elements Mapping}
\label{sec:mapping}

The mapping phase is the first phase in our system. Its purpose is to determine which of the application elements map to the devices involved in the multi-screen experience. 
In any multiscreen scenario, at least two devices are cooperating. The literature refers to the smartTV as the first screen and assigns the expression 'companion screen' \cite{secondscreen3} or 'second screen' \cite{secondscreen2} to a device providing a means of interaction with the smartTV services. 
In this work, the 'principal device' is responsible of processing the main application logic, while the 'secondary device' receives processing results only if it is concerned.
As depicted in Figure \ref{algoArchitecture}, the mapping phase takes as input a query from the application author, when the application is pre-processed offline; or from the user, when the whole process of splitting the application is done at run-time.
In our approach, we have envisaged several possible mapping techniques: based on the analysis of HTML elements, their associated semantics and roles, discussed in Section \ref{sec:semantic}; or based on the visual rendering of elements, discussed in Section \ref{sec:region}. These techniques could be also combined. 
For instance, mapping only interactive elements placed in a certain region of the screen to the secondary device. This phase is extensible and other analysis techniques could be used here. 
The query can therefore be either a simple query indicating which mapping technique should be used along with its specific parameters (e.g., element category or position); or a combined query using boolean logic. The output of the mapping phase is two lists of elements, one for each device.

\subsubsection{Semantic Mapping}
\label{sec:semantic}
As a possible mapping criterion, we describe here a semantic based approach. It is fully automatic, selected by either the author or the user. In the context of multi-screen applications, we analyzed the HTML5 elements defined in the standard to determine their roles and how they could be classified for the purpose of application splitting. We identified four relevant classes: \textbf{interactive} elements (i.e., a, area, button, datalist, form, input, keygen, textarea, nav, optgroup, option, output, select), \textbf{multimedia} elements (i.e., video, audio, source, track), \textbf{non-interactive, non-multimedia~visual} elements (i.e., caption, dialog, figcaption, h1 to h6, hgroup, img, kbd, label, legend, object, p, progress) and \textbf{other} elements. 
As part of the query parameters, the author indicates one or more class of elements to be moved to the secondary device depending on the characteristics of the available devices. 
For instance, if a smartTV is present, the system or the user may decide to move only multimedia elements on that device. As another example, if a touch screen is present, the interactive elements may be moved onto that device. 
This mapping technique produces the lists of elements as follows.
First, for each element of the application, the algorithm compares it to the tag names present in the classes indicated in the author query.
If the element falls within the indicated classes, it is added to the secondary device list.
If the element is a composite element (i.e., div, table, iframe) the algorithm iterates over the its children first.
If there is no match and if the element does not belong to the 'other' class, it is added to the primary device list.
Then, the algorithm detects any change in the element basic role by checking its attributes, mainly declarative event listeners (e.g., 'onclick') since they are responsible of making elements interactive. 
A non-interactive element which role is only displaying content i.e., image, becomes interactive if it has an event listener that lets users interact with it. 
In addition, we exploit the semantic links that are created between HTML elements (e.g., 'for' attribute) by keeping these elements together in the same list. 
\subsubsection{Screen Region Mapping}
\label{sec:region}
We also investigated a region-based approach as a mapping criterion. It is fully automatic but this time based on the application visual rendering. It is selected during run-time by the end-user in her browser. Once the user selects a screen region, the system detects elements within that rectangular region to produce the secondary device list. All other elements are placed in the primary device list.

\subsection{Annotating Elements}
\label{sec:annotating}
As depicted in Figure~\ref{algoArchitecture}, the second phase is the annotation. It prepares the application for the splitting phase and takes as input the lists of elements produced by the mapping phase. 
The annotation algorithm starts by processing each DOM leaf element and sets the value of the 'data-device' attribute to 'device2' for elements in the secondary device list and 'device1' for elements in the primary device list. It should be noted that the previous mapping phase may have populated those lists with only some DOM elements, and not all elements in the tree, for instance only interactive elements or only the elements located in a given region. Thus, each remaining leaf element (resp. parent element) in the DOM tree is annotated with the value of its siblings (resp. its children), if the annotation is the same across siblings (resp. children), or with the value 'dev1\&dev2' if they differ, meaning that the element will be present on both devices. As a result, the application is totally annotated: each element contains metadata information, in a 'data-device' attribute, reflecting its target device(s).

\subsection{Splitting Application Content}
\label{sec:splitting}
After the annotation phase, during run-time, the splitting phase relies on the element metadata information to form two separated applications: a master and a slave application as Figure~\ref{algoArchitecture} shows. 
From the original application, elements annotated with 'device1' or 'dev1\&dev2' values are kept visible on the screen of the primary device. Elements annotated with 'device2' value are hidden on the primary device. This forms the master application. These hidden elements serves as a shortcut whenever the application main logic requires reading or modifying elements of the remote application on the secondary device, thus the term 'virtual splitter'. Elements annotated with 'device2' and 'dev1\&dev2' values are extracted from the original application and imported to the new slave application running on the secondary device. 
On the master application, in addition to the retained original application logic, JS code is added in an instrumentation phase and aims at making the master application capable of working synchronously with its slave (see Section \ref{sec:mirroring}). This code also supports the application dynamism and ensures a dynamic mapping and splitting at run-time. On the slave application, the JS code makes the application capable of collecting user interactions, redirecting them to the master, receiving and integrating changes made to its DOM tree.

\subsection{Mirroring Application Contents}
\label{sec:mirroring}
The virtual splitting phase described in Section~\ref{sec:splitting} duplicates some content between the master and slave applications. The role of the mirroring phase is to ensure that the slave application has a DOM tree that is an accurate mirror of the hidden DOM tree in the master application. It is performed as follows: On the 'primary device', any dynamic change affecting elements of the 'secondary device' (e.g., node modification, removal or creation) is mirrored to that device through change messages. Upon receiving a message, the application running on the 'secondary device' updates its DOM tree and integrates this change. On the 'secondary device', any user interaction (e.g., clicks, data inputs) is captured and propagated to the 'principal device' where the interaction handler is processed.
\begin{figure*}
\centering
\makebox[\textwidth]{
\subfigure[Main Application]{
\includegraphics[height=1.5in,width=0.33\textwidth]{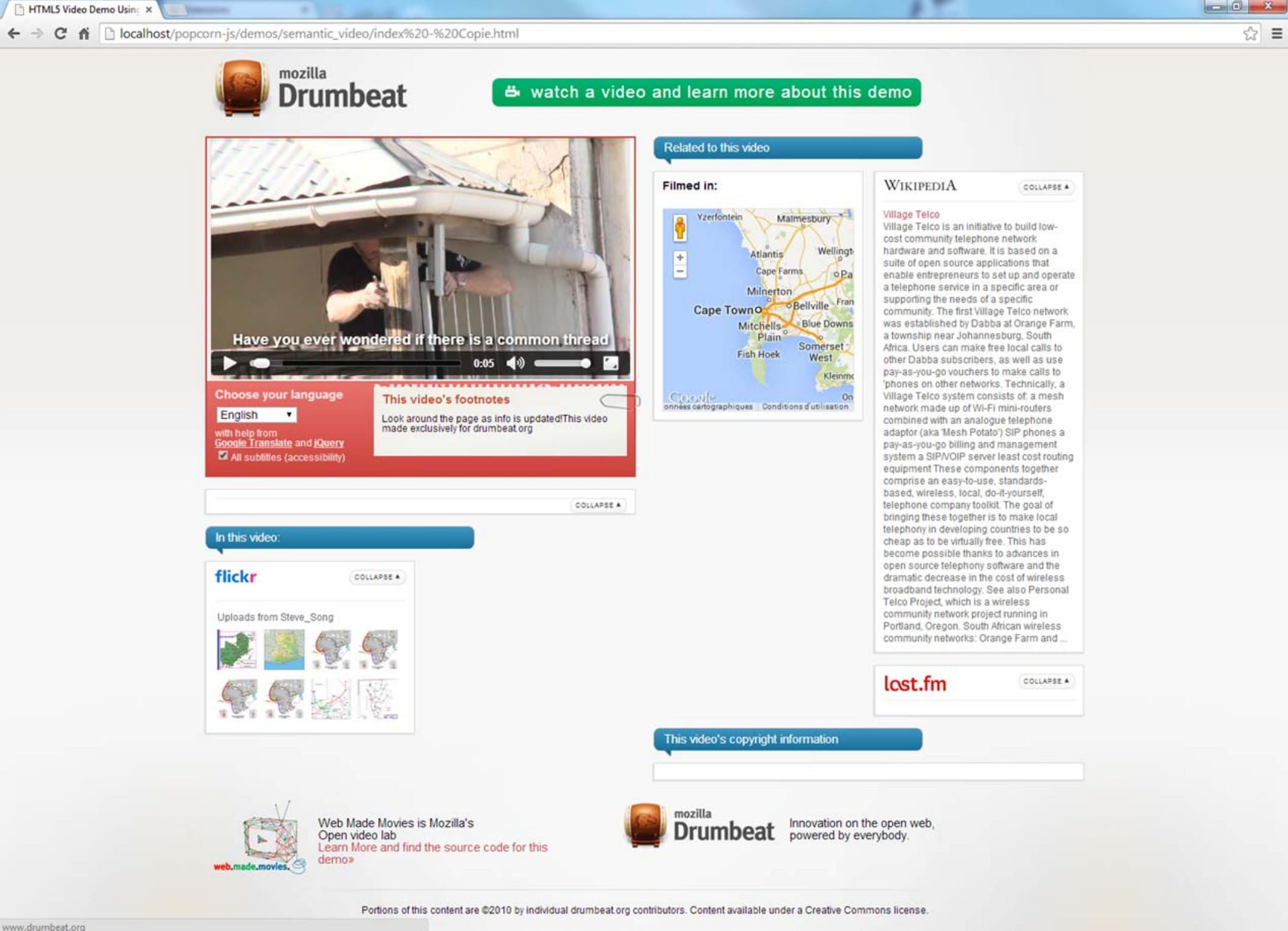}
 \label{sf:0}
}
\subfigure[Master Application]{
\includegraphics[height=1.5in,width=0.33\textwidth]{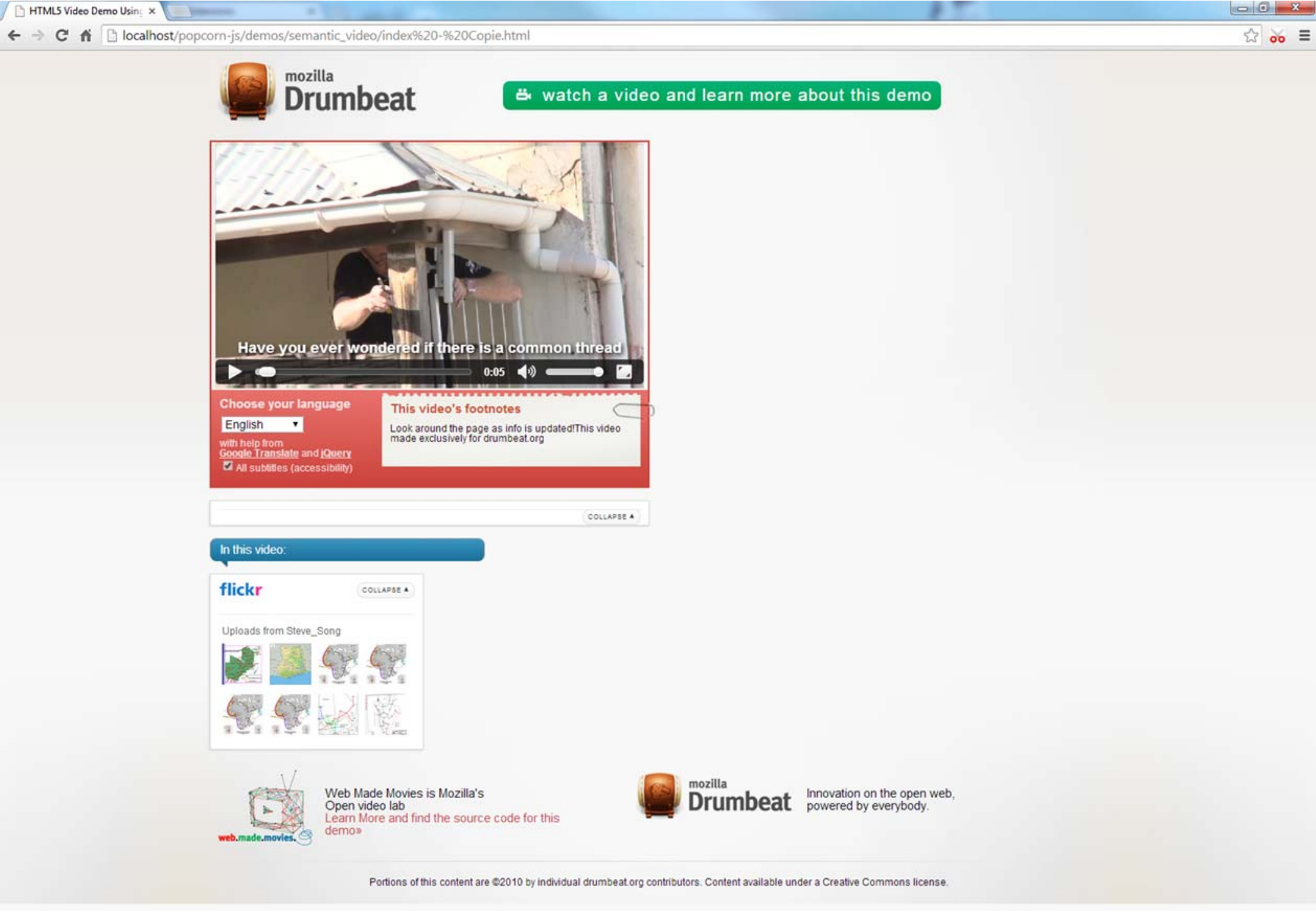}
 \label{sf:1}
}
\quad
\subfigure[Slave Application]
{\includegraphics[height=1.5in,width=0.33\textwidth]{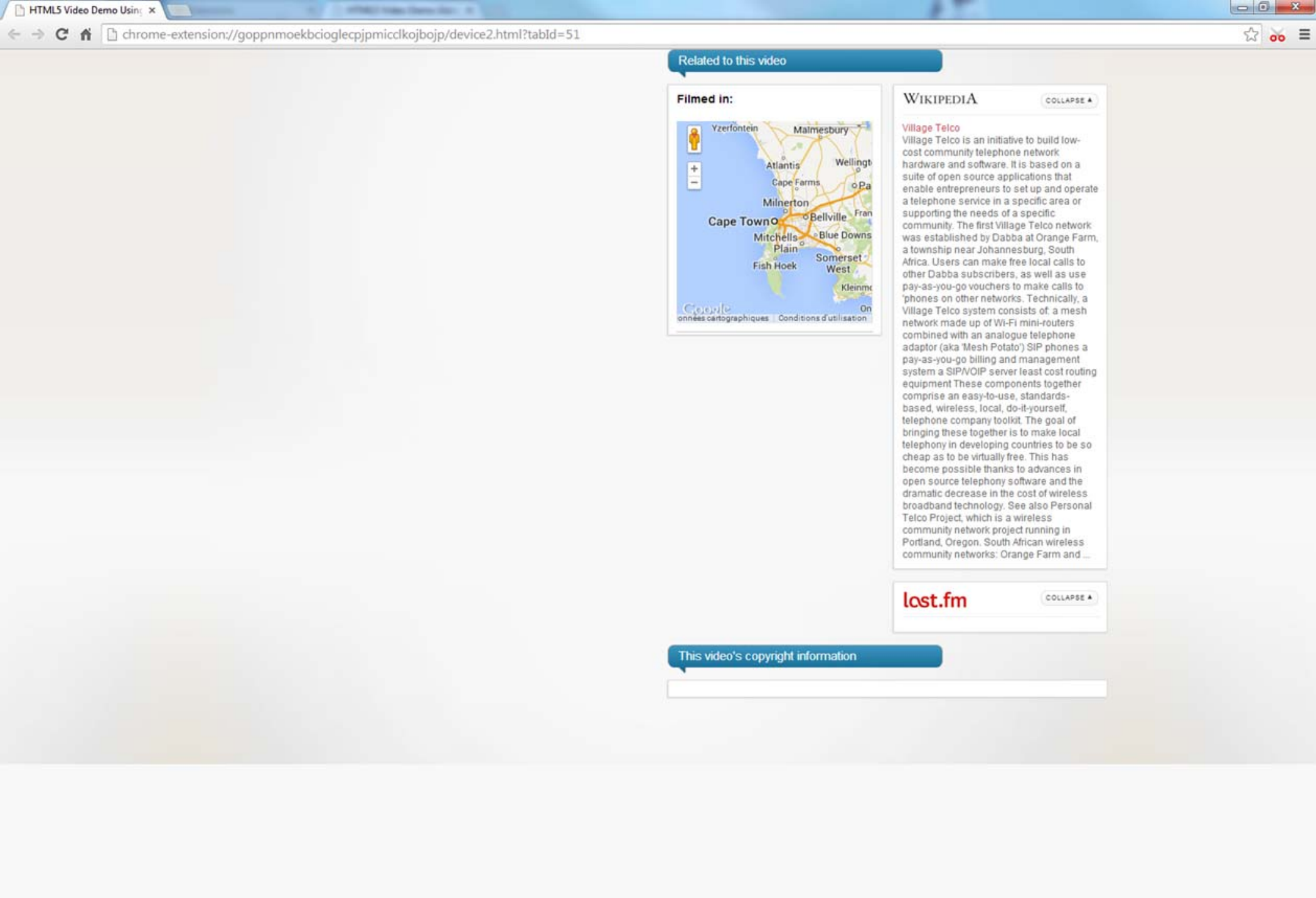}
 \label{sf:2}
}
}
\caption{Splitting The Semantic Video Application Based On The Screen-Region Criterion}
\label{fig:masterapp}
\end{figure*}

\section{Implementation and results}
\label{sec:implementation}
\subsection{Implementation}
\label{subsec:setup}
We decided to implement the virtual splitter as a Google Chrome extension, 
first to enable on-the-fly instrumentation of the application without having to change the application itself,
and second for the better debugging environment compared to the situation on devices. 
Master and slave applications are rendered as tabs in the browser and communication between them is done 
using the postMessage API that is similar to the communication mecanism in COLTRAM\cite{coltramEuroITV}.

As discussed in Section \ref{sec:mirroring}, to detect relevant changes in the DOM we use the Mutation Summary library\footnote{see \url{http://code.google.com/p/mutation-summary}} which is based on the working draft of the Mutation Observer API \footnote{see \url{http://www.w3.org/TR/domcore}}. A Mutation-Summary object is configured to watch changes made to elements with 'device2' annotation. If any change happens to these elements, their descendants or attributes, the extension sends a message to the 'secondary device'. The message contains a list of changes. Each change object consists of the type of change, the concerned node, its position in the DOM tree (i.e., parent and previous sibling node), the concerned attribute(s) and the new value of a text node.
However, the Mutation Summary library suffers from some limitations. For instance, it cannot detect changes made to HTML elements using JS functions especially if they are not reflected on the DOM tree. 
To overcome this limitation, as well as supporting dynamism, we use the Monkey Patching technique \cite{monkey-patching}, to extend in JS some native browser functions with custom code, in particular: the 'createElement' function is extended to detect the creation of new elements and to trigger dynamically the mapping, annotation and splitting of these elements; the 'setAttribute' function to update the Mutation Summary configuration, to enable the mirroring of newly created attributes; and the 'addEventListener' function to overcome the limitation of the Mutation Summary library, and to replace an event handler triggered on the slave application to a call to the master application.

\subsection{Results and Discussion}
\label{subsec:results}
We tested our system on different applications from simple static pages to dynamic applications, among them: a semantic video application\footnote{see \url{http://popcornjs.org/demo/semantic-video}}, relying on 
the Popcorn and JQuery libraries and showing various information (e.g., map, text, images) synchronized with a video. 
We first used the region-based mapping. On the video application, we separated the video and Flick'r images from the 
additional information as Figure~\ref{fig:masterapp} shows. This experiment verified the performance of the mirroring 
phase by maintaining a reliable mirror and the synchronization between both master and slave applications.
In addition, no compatibility issues were reported between our instrumented code and the JS libraries.
We then used the semantic mapping to split a YouTube page and to separate the interactive class of elements from the other classes (i.e., non-interactive and multimedia). As a result, the video runs on the master with all the comments of users while all buttons, anchors, guide container that proposes additional videos to watch later are moved to the secondary device.

Based on this, we identified a few areas for future improvement. This includes mainly the re-organization of application layout based on devices screen characteristics, the solving of some problems related to the use of relative URLs, the handing of HTML elements such as Canvas that have no inner DOM representation but are controlled by JS code. We will also conduct more systematic testing and validation.

\section{Perspectives and Conclusions}
In the multi-screen context, this paper proposed a system to transform existing applications from single-screen to multi-screen applications based on author or user choices. 
The system consists of an automatic and extensible mapping phase that analyzes the application semantically, visually or a combination of these two, an annotation and a virtual splitting phase that result in master-slave applications. A mirroring phase ensures the correct functionality and synchronization between both parts and a dynamic splitting process. 
We validated our system on two existing applications: YouTube and semantic-video, and we verified the correct content mapping, synchronization and application functionality.
As a future step, we aim at implementing our system in the COLTRAM multi-screen platform and extending this system with a context driven splitting technique that collects information concerning involved devices and creates dynamically adaptive mapping criteria.
\label{sec:conclusion}
\bibliographystyle{plain}
\bibliography{docEng}
\end{document}